\documentclass[pra,aps,twocolumn,showpacs,tightenlines]{revtex4}
\usepackage{graphics,bm}
\usepackage{graphicx}
\newcommand{\beq}{\begin{equation}}
\newcommand{\eeq}{\end{equation}}
\newcommand{\bqa}{\begin{eqnarray}}
\newcommand{\eqa}{\end{eqnarray}}
\protect

\begin{document}

\title{Dimensional and Temperature Crossover in Trapped Bose Gases}
\author{U. Al Khawaja$,^1$\footnote{Present Address: Physics Department,
United Arab Emirates University, P.O. Box 17551, Al-Ain, United
Arab Emirates.} N.P. Proukakis,$^{2}$ J.O.
Andersen$,^1$\footnote{Present Address: Nordita, Blegdamsvej 17,
2100 Copenhagen, Denmark. } M. W. J. Romans,$^1$ and H.T.C.
Stoof$^1$} \affiliation{ \it $^1$Institute for Theoretical
Physics, University of Utrecht, Leuvenlaan 4, 3584 CE Utrecht, The
Netherlands
\\$^{2}$Department of Physics, University of Durham,
South Road, Durham DH1 3LE, United Kingdom.}

\date{\today}

\begin{abstract}
We investigate the long-range phase coherence of homogeneous and
trapped Bose gases as a function of the geometry of the trap, the
temperature, and the mean-field interactions in the weakly
interacting limit. We explicitly take into account the
(quasi)condensate depletion due to quantum and thermal
fluctuations, i.e., we include the effects of both phase and
density fluctuations. In particular, we determine the phase
diagram of the gas by calculating the off-diagonal one-particle
density matrix and discuss the various crossovers that occur in
this phase diagram and the feasibility of their experimental
observation in trapped Bose gases.
\end{abstract}


\maketitle

\section{Introduction}
\label{into}

The weakly-interacting Bose gas in three dimensions has been
studied in great detail over the past 50 years~\cite{grif}. Below
a critical temperature $T_c\simeq({2\pi\hbar^2/mk_B})
\left[n/\zeta(3/2)\right]^{2/3}$, where $n$ is the total density,
the gas is in a Bose-Einstein condensed state. As with almost all
thermodynamic phase transitions in three dimensions, a
Bose-Einstein condensate leads to long-range order in the system.
In this particular case, the long-range order is determined by the
behavior of the off-diagonal one-particle density matrix
$\langle\psi^{\dagger}({\bf x})\psi({\bf 0})\rangle$. If the
one-particle density matrix goes to a constant $n_c$ in the limit
$|{\bf x}|\rightarrow\infty$, a Bose-Einstein condensate is present
and $n_c$ is the condensate density.

The physics of one and two-dimensional Bose gases is very
different from that of the three-dimensional one, which makes
these low-dimensional systems very interesting. From a theoretical
point of view, this difference is caused by the enhanced
importance of the phase fluctuations~\cite{mullin,jason,1d,2d}. In
fact, the phase fluctuations are so large that in a homogeneous
one-dimensional Bose gas at all temperatures and in a homogeneous
two-dimensional Bose gas at any nonzero temperature, Bose-Einstein
condensation cannot take place. This is the content of the
Mermin-Wagner-Hohenberg theorem~\cite{mermin,hohen}.

A natural question then arises: What happens if we have a large
three-dimensional box with a Bose-Einstein condensed gas and
squeeze two of the sides so as to obtain a one-dimensional system?
Similarly, what happens if we take the same box containing a
Bose-Einstein condensed gas at a nonzero temperature and squeeze
one of the sides so that a two-dimensional system results? These
are questions that we would like to address in the present paper.
Until recently, these questions could not be discussed on the
basis of a microscopic theory, since no accurate equation of state
existed for one and two-dimensional Bose gases. However, we have
recently developed an improved many-body T-matrix theory for
partially Bose-Einstein condensed atomic gases by treating the
phase fluctuations exactly~\cite{jh,jhun}. This mean-field theory
is valid in arbitrary dimensions and accounts also for the
(quasi)condensate depletion. It is therefore very suitable for
describing the crossovers mentioned above.

Low-dimensional Bose gases are presently receiving a large amount
of attention due to the recent experimental realization of one-
and two-dimensional condensates in
traps~\cite{exp100,lowdketterle} as well as one-dimensional gases
on microchips~\cite{{exp200},{exp500},{exp300}}. Because these
experiments deal with trapped, and therefore inhomogeneous, Bose
gases it is important to consider this situation as well. The
Mermin-Wagner-Hohenberg theorem~\cite{mermin,hohen} theorem is
valid only in the thermodynamic limit, and so does not immediately
apply to trapped Bose gases. In the trapped case, it turns out
that under certain conditions the phase is coherent over a
distance of the order of the size of the system and a ``true''
condensate is present. Under other conditions, the phase is
coherent over a distance less than the size of the system and only
a so-called ``quasicondensate''~\cite{popov} is
present~\cite{mullin,jason,2d,1d}. In fact, this situation can
even occur in elongated three-dimensional Bose gases
\cite{{petrov22},new,orsay1,{aspect},hannover1,walraven1}. As a
result, similar questions as above arise: What happens to the
phase fluctuations in a one-dimensional trapped Bose gas as we
vary the temperature and the trap frequency? This is another
question that we consider here.

The paper is organized as follows. In Sec.~\ref{sec1}, we briefly
discuss the modified Popov theory presented in
Refs.~\cite{jh,jhun}. In Sec.~\ref{sec2}, we discuss dimensional
crossovers and finite-size effects in a homogeneous Bose gas. In
Sec.~\ref{sec3}, the temperature crossover in both homogeneous and
trapped Bose gases in one dimension is examined. Finally, we
summarize and conclude in Sec.~\ref{conc}.

\begin{figure}[htb]
\begin{center}
\includegraphics[width=8cm]{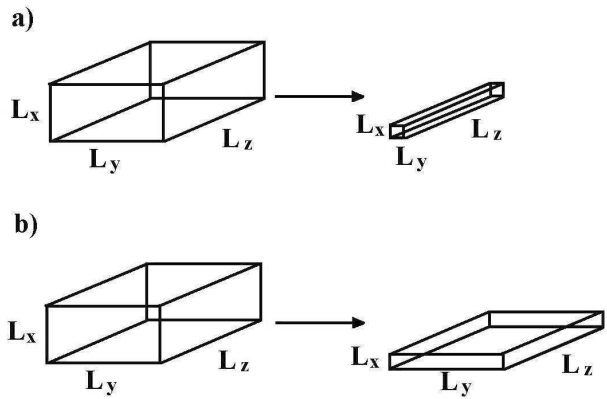}
\end{center}
\caption[a]{
Illustration of the geometry for a) the dimensional crossover from
a three-dimensional to a one-dimensional Bose gas, and b) the
dimensional crossover from a three-dimensional to a
two-dimensional Bose gas.} \label{fig1}
\end{figure}

\section{Modified Popov Theory}
\label{sec1}

It is well known that the usual Popov theory for partially
Bose-Einstein condensed gases suffers from infrared divergences in
the equation of state at all temperatures in one dimension and at
any nonzero temperature in two dimensions. These infrared problems
arise because only quadratic fluctuations around the mean field
have been taken into account. Specifically, the annihilation
operator for the field is written as $\hat{\psi}({\bf
x})=\sqrt{n_0}+\hat{\psi}^{\prime}({\bf x})$ and terms in the
Hamiltonian that are of third order or higher in the fluctuating
field $\hat{\psi}^{\prime}({\bf x})$ are neglected. In
Refs.~\cite{jh,jhun}, it was shown that the problem of infrared
divergences in the equation of state can be solved by taking into
account phase fluctuations exactly and not only up to second
order. The quadratic contribution to the density from the phase
fluctuations is $n_0\langle\hat{\chi}({\bf x})\hat{\chi}({\bf
x})\rangle$, while an exact result would give no contribution at
all. This follows from the identity $n_0\langle
e^{-i\hat{\chi}({\bf x})} e^{i\hat{\chi}({\bf x})} \rangle =
n_0\left(1 + \langle \hat{\chi}({\bf x})\hat{\chi}({\bf x})
\rangle + \dots\right) =1$. In order to obtain the correct result,
we must therefore subtract the quadratic contribution from the
phase fluctuations. This ultimately results in the following
expressions for the density $n$ and chemical potential $\mu$
\begin{eqnarray}\nonumber
n&=&n_0+{1\over V}\sum_{\bf k}
\Bigg[
{\epsilon_{\bf k}-\hbar\omega_{\bf k}\over2\hbar\omega_{\bf k}}
+{n_0T^{\rm 2B}(-2\mu)\over2\epsilon_{\bf k}+2\mu}
\\&&
+{\epsilon_{\bf k}\over\hbar\omega_{\bf k}} N(\hbar\omega_{\bf k})
\Bigg] \;, \label{h1} \\
\mu&=&(2n-n_0)T^{\rm 2B}(-2\mu) \label{h2} =(2n'+n_0)T^{\rm
2B}(-2\mu) \;,
\end{eqnarray}
where $n_0$ is the (quasi)condensate density and $n'=n-n_0$
represents the depletion of the (quasi)condensate due to quantum
and thermal fluctuations. It is important to point out that this
theory takes the (quasi)condensate depletion into account, and can
thus work even in the limit of $n_0/n \ll 1$, as will become
apparent in subsequent sections. The Bogoliubov quasiparticle
dispersion relation is given by $\hbar\omega_{\bf k}
=\left[\epsilon_{\bf k}^2+2n_0T^{\rm 2B}(-2\mu)\epsilon_{\bf
k}\right]^{1/2}$, where $\epsilon_{\bf k}=\hbar^2{\bf k}^2/2m$.
Note also that the energy argument of the $T$-matrix is $-2\mu$,
because this is precisely the energy it costs to excite two atoms
from the condensate~\cite{schick,fisher}.

The Bose distribution function is given by
$N(\varepsilon)=1/[\exp(\varepsilon/k_BT)-1]$, where $k_B$ is the
Boltzmann constant and $T$ is the temperature. In addition,
$T^{\rm 2B}$ is the two-body $T$-matrix and $V$ is the volume of
the system. The two-body $T$ matrix takes into account successive
two-particle scattering in the vacuum, but neglects many-body
effects of the medium. These many-body effects can be taken into
account by using the many-body $T$ matrix instead of the two-body
$T$-matrix~\cite{jhun}. For simplicity we shall use the two-body
$T$-matrix for the most part of this work. However, in Sec. IV B
we will actually use the many-body T-matrices to make the
calculations for the trapped case as accurate as possible.

It is important to realize that the momentum sum on the right-hand
side of Eq.~(\ref{h1}) includes the term with ${\bf k}={\bf 0}$.
This is a result of the fact that if we include fluctuations we
have that $\langle\hat{\psi}_{\bf 0}^{\dagger}\hat{\psi}_{\bf
0}\rangle$, where ${\hat\psi}_{\bf 0}=\int{d{\bf x
}({\hat\psi}({\bf x})/V)}$, is not exactly equal to $n_0$. This
difference is most important at nonzero temperatures and by taking
the limit ${\bf k}\rightarrow{\bf 0}$ in Eq.~(\ref{h1}), we find
that the contribution from the zero-momentum state to the density
equals $n_0+(1/V)(k_BT/2n_0T^{2B})$. This result can be physically
understood by realizing that $\langle\hat{\psi}_{\bf
0}^{\dagger}\hat{\psi}_{\bf 0}\rangle$ can be found by calculating
the average of $|\psi_0|^2$ with a probability distribution that
obeys \bqa P(|\psi_0|)=e^{-{VT^{2B}\over k_BT}
\left(-n_0|\psi_0|^2+{1\over2}|\psi_0|^4\right)}\;. \eqa To be
consistent, the above calculation has to be performed in the
Bogoliubov approximation as well.

We define the normalized  first-order
correlation function $g({\bf x})$ by
\begin{equation}
\langle{\hat\psi}^\dagger({\bf x}){\hat\psi({\bf0})}\rangle=
\sqrt{n_0({\bf x})n_0({\bf 0})}\;g({\bf x})\;.
\label{nfi}
\end{equation}
In the large-$|{\bf x}|$ limit, it thus takes the form
\begin{equation}
g({\bf x})\simeq e^{-{1\over2}\langle\left[\hat{\chi}({\bf
x})-\hat{\chi}({\bf 0})\right]^2 \rangle}\;. \label{d11}
\end{equation}
For a homogeneous Bose gas, the phase fluctuations are determined
by~\cite{jh,jhun} \bqa\nonumber \langle\left[\hat{\chi}({\bf
x})-\hat{\chi}({\bf 0})\right]^2\rangle &=&{T^{\rm 2B}(-2\mu)\over
V}\sum_{\bf k}\Bigg[ {1\over\hbar\omega_{\bf k}}
\left[1+2N(\hbar\omega_{\bf k})\right]
\\&&
-{1\over\epsilon_{\bf k}+\mu}\Bigg] \left[1-\cos({\bf
k}\!\cdot\!{\bf x})\right] \;. \label{usa1} \eqa The analogous
expression for the one-dimensional harmonically confined
condensate has also been calculated. In the Thomas-Fermi limit and
neglecting quantum fluctuations, it reads \cite{jhun}
\begin{eqnarray}\nonumber
&&\langle \left[ \hat{\chi}(z)-\hat{\chi}(0) \right]^2 \rangle
={4\pi \kappa l_z^4\over R_{\rm TF}^3} \sum_{j=0}N(\hbar\omega_j)
\\ \nonumber
&\times&\Bigg[ A_j^2\left(P_j(z/R_{\rm TF})-P_j(0)\right)^2
\\
&-&B_j^2\left({P_j(z/R_{\rm TF})\over 1-(z/R_{\rm
TF})^2}-P_j(0)\right)^2 \Bigg]\;, \label{chifinal}
\end{eqnarray}
where $A_j=\sqrt{(j+1/2)\mu^\prime/\hbar\omega_j}$,
$B_j=\sqrt{(j+1/2)\hbar\omega_j/\mu^\prime}$, and $P_j(z)$ is the
Legendre polynomial of order $j$. Here $\mu^\prime=\mu-2\kappa
n^\prime(0)$, $n^\prime(0)$ is depletion in the center of the
trap, and the coupling constant $\kappa=a/2\pi\ell_\perp^2$
results from averaging the three-dimensional two-body $T$-matrix, which is
equal to $4\pi a \hbar^2/m$ with $a$ the $s$-wave scattering
length, over the cross-sectional area of the one-dimensional
condensate. Furthermore, $\ell_\perp$ is the harmonic oscillator
length in the radial direction. In the axial direction, the trap
length is $\ell_z=\sqrt{\hbar/m\omega_z}$ where $\omega_z$ is the
harmonic oscillator frequency in the $z$-direction. The quantity
$R_{\rm TF}$ denotes the temperature-dependent Thomas-Fermi
radius, defined by the point at which the quasicondensate density
$n_0(z)=(\mu^\prime-V^{\rm trap}(z))$ vanishes. Here $V^{\rm
trap}(z)$ is the harmonic trapping potential in the axial
direction. The frequencies $\omega_j$ are given by
$\omega_j=\sqrt{j(j+1)/2}\;\omega_z$.

If the function $g({\bf x})$ aproaches a constant in the
large-$|{\bf x}|$ limit, the system contains a true Bose-Einstein
condensate. If $g({\bf x})$ goes to zero algebraically in the
limit $|{\bf x}|\rightarrow\infty$, the system is said to contain a
quasicondensate. Finally, if $g({\bf x})$ goes to zero
exponentially fast, the system is in the normal state. In the
following, we calculate the correlation function $g({\bf x})$ for
various situations.

\section{Dimensional Crossover in a Homogeneous Bose Gas}
\label{sec2} In this section, we investigate the dimensional
crossover from three to one dimension at zero temperature, as well
as the dimensional crossover from three to two dimensions at
nonzero temperature. The relevant geometry is illustrated in
Fig.~\ref{fig1}.

\subsection{Crossover from three dimensions to one dimension at zero temperature}
We consider a uniform Bose gas in a box with lengths $L_x$, $L_y$,
and $L_z$ and impose
periodic boundary conditions.
Due to the boundary conditions, the three components of the wave vector
$\bf k$ take on discrete values, i.e.,
$k_i=2\pi n_i/L_i$, where $i=x$, $y$, $z$ and $n_i$ is an integer.
Substituting this into Eq.~(\ref{usa1}), we obtain
\begin{eqnarray}\nonumber
\langle \left[\hat{\chi}(z)-
             \hat{\chi}({0}) \right]^2 \rangle&=&
 {1\over2N_0}\sum_{n_x,n_y,n_z}
\left[{1\over\sqrt{\left(k\xi\right)^4
+\left(k\xi\right)^2}}\right.
\\ &&
\left.-{1\over \left(k\xi\right)^2+{1\over2}}\right]
\left[1-\cos\left({n_z z\over \lambda_z\xi}\right)\right]\;.
 \label{gen1}
\end{eqnarray}
where $N_0$ is the number of atoms in the (quasi)condensate,
$\lambda_z=L_z/2\pi\xi$, and
$\xi=\hbar/[4mn_0T^{\rm 2B}(-2\mu)]^{1/2}$ is the correlation
length. Note that we have neglected the contribution from
$n^{\prime}$ to the chemical potential, which is a good
approximation in the weak-coupling limit we are considering. Note
that the argument of the cosine has been simplified by choosing
the direction of $\bf x$ to be along the $z$-axis.

For very elongated systems, one side of the box, is much larger
than the other two, and we choose $L_z\gg L_x,L_y$. If we
also take $L_z$ to be much larger than the correlation length, we
can approximate the sum over $n_z$ by an integral. In the limit
$z/\xi\gg1$, we obtain
\begin{eqnarray}
\langle \left[\hat{\chi}(z)-
             \hat{\chi}({0}) \right]^2 \rangle&
=& {\lambda_z\over N_0}
\Bigg[\gamma-{\pi\over\sqrt{2}}+\ln{2}+\ln{\left({z\over\xi}\right)}
\nonumber\\
&&\hspace{-2cm}
+\sum_{n_x,n_y}\!\!^{\prime}\left[
{K_0\left(-{1\over t_{x,y}}\right)\over\sqrt{t_{x,y}}}
-{\pi\over2\sqrt{t_{x,y}}+{1\over2}}
\right]\Bigg]\;, \label{gen2}
\end{eqnarray}
where \bqa t_{x,y}=
\left({n_x\over\lambda_x}\right)^2+\left({n_y\over\lambda_y}\right)^2
\eqa and the prime on the sum indicates that the term where
$n_x=n_y=0$ is omitted. Here $K_0(z)$ is a modified Bessel
function of the second kind, $\lambda_{x,y}=L_{x,y}/2\pi\xi$, and
$\gamma\simeq 0.5772$ is Euler's constant.

The one-dimensional limit is obtained by letting
$L_{x,y}\rightarrow0$. This is equivalent to keeping only the
first term in the right-hand side of Eq.~(\ref{gen2}), thus
\bqa\nonumber \langle \left[\hat{\chi}(z)-
             \hat{\chi}({0}) \right]^2 \rangle
&=&
             {1\over2\pi{n_0}\xi}
\left[\gamma-{\pi\over\sqrt{2}}+\ln2\right.
\\
&&\left.
+\ln{\left({z\over\xi}\right)}\right]\;,
             \label{gen3}
\eqa where $n_0=N_0/L_z$ is the one-dimensional quasicondensate
density. The result~(\ref{gen3}) shows that the correlation
function $g(z)$ at large distances falls off algebraically with
the exponent $\eta=1/4\pi n_0\xi$. In the weakly-interacting limit
$4\pi n\xi\gg1$ the depletion is small, so that we indeed have
that $n_0\simeq n$. Keeping this in mind, Eq.~(\ref{gen3}) is in
complete agreement with the exact result obtained by
Haldane~\cite{haldane}.

The three-dimensional limit it obtained by letting $L_{x,y}\rightarrow\infty$.
The discrete sums in Eq.~(\ref{gen2}) then become integrals.
Performing the integrations, we obtain in the limit $z/\xi\rightarrow\infty$
\begin{equation}
\langle \left[\hat{\chi}(z)-
             \hat{\chi}({0}) \right]^2 \rangle=
             {\sqrt{2}\pi-4\over16\pi^2\;{n_0\xi^3}}
             \label{gen4}\;,
\end{equation}
where $n_0=N_0/L_xL_yL_z$ is now the three-dimensional
quasicondensate density. Eq.~(\ref{gen4}) shows that the
correlation function in three dimensions goes to a constant, so we
have a ``true'' condensate. Note, however, that in three
dimensions the exponent in Eq.~(\ref{d11}) does not vanish and
therefore that $n_0$ appearing in Eqs.~(\ref{h1}) and (\ref{h2})
is not exactly equal to the condensate density $n_c$. In fact,
Eq.~(\ref{gen4}) shows that the condensate density $n_c$ is given
by \bqa\nonumber
n_c&=&n_0\exp\left({4-\sqrt{2}\pi\over32\pi^2\;{n_0\xi^3}}\right)\\
&=&n\left[1-{8\over3}\sqrt{{na^3}\over\pi}+O(na^3)\right]\;, \eqa
where $n$ is the total density of the gas. This result is in full
agreement with that of the Popov or Bogoliubov theory.

The dimensional crossover behavior can be investigated in detail from
Eq.~(\ref{gen2}) by calculating the phase fluctuations $\langle
\left[\hat{\chi}(z)-\hat{\chi}({0}) \right]^2 \rangle$ for
different finite values of $\lambda_x$ and $\lambda_y$. Comparing
Eqs.~(\ref{gen3}) and~(\ref{gen4}), it follows that the phase
fluctuations scale with the product $\lambda_x\lambda_y$ in the
three-dimensional limit. In order to plot the phase fluctuations
for different sizes of the box, it is therefore convenient to
multiply $\langle \left[\hat{\chi}(z)-\hat{\chi}({0})
\right]^2\rangle$ by a scaling function $f(\lambda_x,\lambda_y)$
that goes to $1/\lambda_x\lambda_y$ in the three-dimensional
limit. Moreover, it follows from Eq~(\ref{gen3}), that
$f(\lambda_x,\lambda_y)$ must approach unity in the one-dimensinal
limit. A simple choice for a scaling function that has these
properties is \bqa
f(\lambda_x,\lambda_y)&=&{1\over(\lambda_x+1)(\lambda_y+1)}\;.
\eqa In the following, we take $\lambda_x=\lambda_y=\lambda$ for
simplicity.

\begin{figure}[htb]
\begin{center}
\includegraphics[width=8cm]{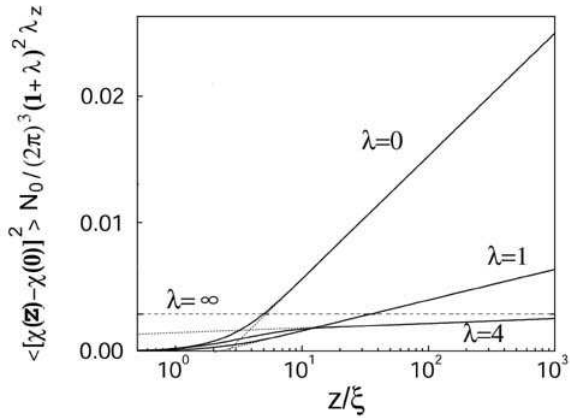}
\end{center}
\caption[a]{
Zero-temperature crossover behavior in the phase fluctuations of a
three-dimensional homogeneous Bose-Einstein condensate. The three
full lines show the phase fluctuations $\langle
\left[\hat{\chi}(z)-\hat{\chi}({0}) \right]^2 \rangle N_0
/(2\pi)^3(1+\lambda)^2\lambda_z$ as a function of $\ln{(z/\xi)}$,
for three values of $\lambda$. The dotted lines are calculated from
Eq.~(\ref{gen2}). } \label{fig2}
\end{figure}

In Fig.~\ref{fig2}, we plot $\langle
\left[\hat{\chi}(z)-\hat{\chi}({0}) \right]^2 \rangle N_0
/(2\pi)^3(1+\lambda)^2\lambda_z$ as a function of $\ln{(z/\xi)}$
for three values of $\lambda$. In the three-dimensional limit
$\lambda\rightarrow\infty$, the phase fluctuations are independent of
$z$ for large $z/\xi$. For finite values of $\lambda$, the curves
deviate from the $\ln{(z/\xi)}$ behavior due to contributions from
the sum in Eq~(\ref{gen2}). It should be noted that convergence of
the sum in the left-hand side of Eq.~(\ref{gen1}) depends on the
value of $\lambda$. For large $\lambda$, we included more terms in
the sum to obtain convergence. The summations were terminated when
the error is less than 0.1$\%$ of the three-dimensional
result.

\subsection{Crossover from three dimensions to two dimensions at a
nonzero temperature} The nonzero-temperature part of the phase
fluctuations is given by \bqa
{1\over N_0}\sum_{n_x,n_y,n_z}{
1-\cos\left({{n_zz\over\lambda_z\xi}}\right) \over
k\xi\sqrt{(k\xi)^2+1} \left(e^{2n_0T^{\rm
2B}k\xi\sqrt{(k\xi)^2+1}/k_BT}-1\right)}\;. \nonumber\\
\label{phaset} \eqa The first step consists of taking the limit
$L_y=L_z\rightarrow\infty$. The sums over $n_y$ and $n_z$ then
become integrals. The two-dimensional limit is obtained by letting
$L_x\rightarrow0$, which is equivalent to keeping only the $n_x=0$
term in the above sum. After going to polar coordinates, this
gives \bqa\nonumber &&{\lambda^2\over N_0}
\int_{0}^{\infty}\int_{0}^{2\pi} {d(k\xi)d\phi
\over\sqrt{(k\xi)^2+1}\left(e^{2n_0T^{\rm
2B}k\xi\sqrt{(k\xi)^2+1}/k_BT}-1\right)}
\\ \nonumber
&& \times \left[1-\cos\left({kz}\cos\phi\right)\right] \eqa with
$\lambda_y=\lambda_x=\lambda$. Integration over $\phi$ leads to
\bqa\nonumber &&{2\pi\lambda^2\over N_0}\int_{0}^{\infty} d(k\xi){
1-J_0(kz)\over\sqrt{(k\xi)^2+1}\left(e^{2n_0T^{\rm
2B}k\xi\sqrt{(k\xi)^2+1}/k_BT}-1\right)}\;, \eqa where $J_0(z)$ is
a Bessel function of the first kind. The integral behaves as
$k_BT\ln(z/\xi)/2n_0T^{\rm 2B}$ in the limit $z/\xi\gg1$. Thus the
first-order correlation function obeys \bqa\nonumber
g(z)\simeq{1\over({z/\xi})^{1/n_0\Lambda^2}}\;, \eqa where
$\Lambda=\sqrt{2\pi\hbar^2/mk_BT}$ is the thermal de Broglie
wavelength.

The three-dimensional limit is obtained by letting
$L_x=L_y=L_z\rightarrow\infty$ in Eq.~(\ref{phaset}). The sums
over $n_x$, $n_y$, and $n_z$ then become integrals. Going to polar
coordinates and integrating over angles, we obtain \bqa\nonumber
&&\frac{2 \pi \lambda^3}{N_0}\int_{0}^{\infty}
{d(k\xi)\;k\xi\over\sqrt{(k\xi)^2+1}} {1\over(e^{2n_0T^{\rm
2B}k\xi\sqrt{(k\xi)^2+1}/k_BT}-1)}
\\\nonumber
&&\left[1-{\sin\left({kz}\right)\over kz}\right]\;. \label{3df}
\eqa The first term is a constant, independent of $z$. The second
term is oscillating and goes to zero in the limit
$z/\xi\rightarrow\infty$. This shows that the total integral, and
thus the phase fluctuations, go to a nonzero constant and that
Bose-Einstein condensation is possible in three dimensions at
nonzero temperature. For low temperatures, where the phonons give
the main contribution to the above integral, this constant is
proportional to $1/n_0\xi\Lambda^2 \ll 1$. Combining, the above result
we see that the crossover behavior is therefore qualitatively very
similar to the one presented in Fig.~\ref{fig2}.

\section{Temperature Crossover in a finite one-dimensional Bose gas}
\label{sec3} In this section, we first study the temperature
crossover in a homogeneous one-dimensional Bose gas. The aim is to
capture the essential crossover physics and to find the relevant
parameters. Having established the relevant parameters, we examine
a harmonically trapped one-dimensional Bose gas.

\subsection{Homogeneous Bose gas}
In this subsection we calculate the nonzero-temperature
correlation function of a homogeneous Bose gas in one dimension.
We construct the phase diagram by plotting this function versus
the temperature $T$ and the length $L$ of the system.
In detail the procedure is as follows.

We first solve the two coupled Eqs.~(\ref{h1}) and (\ref{h2}) for
the (quasi)condensate density $n_0$ at a fixed temperature, system
size $L$, and for a total density $n=N/L$, where $N$ is the fixed
total number of atoms. Then we use this solution in
Eq.~(\ref{usa1}) to calculate the correlation function. To obtain
a universal phase diagram, we need to scale $T$ and $L$ to some
characteristic temperature and length in the system. In the
present case our homogeneous system is characterized by its
correlation length $\xi$. The correlation length is normally
defined in terms of the condensate density. However, above the
Bose-Einstein transition temperature, the normal state cannot be
characterized by this length. Therefore we define the correlation
length in terms of the total density, rather than the condensate
density, namely $\xi=1/\sqrt{8\pi\kappa n}$. As a result, we scale
lengths to $\xi$ and temperature to $T_0=\hbar^2/8\pi m\xi^2$. The
scaled temperature $T/T_0$ is equivalent to $(\xi/\Lambda)^2$,
where $\Lambda=\sqrt{2\pi\hbar^2/ m k_{\rm B}T}$ is the de Broglie
wavelength.

\begin{figure}[htb]
\begin{center}
\includegraphics[width=8cm]{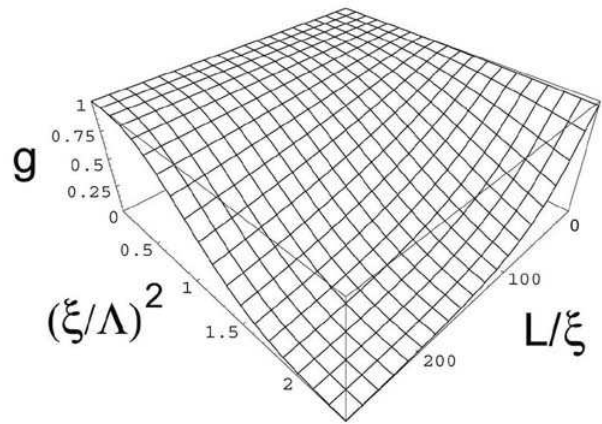}
\end{center}
\caption[a]{\narrowtext The correlation function of a homogeneous
one-dimensional condensate as a function of $L/\xi$ and
$(\xi/\Lambda)^2$. The correlation function is calculated at
$z=0.9L$ and for a total number of atoms $N=2000$. } \label{fig3}
\end{figure}

In Fig.~\ref{fig3} we plot the correlation function at $z=0.9L$
versus the dimensionless variables $L/\xi$ and $(\xi/\Lambda)^2$
for a fixed number of atoms.
\begin{figure}[htb]
\begin{center}
\includegraphics[width=8cm]{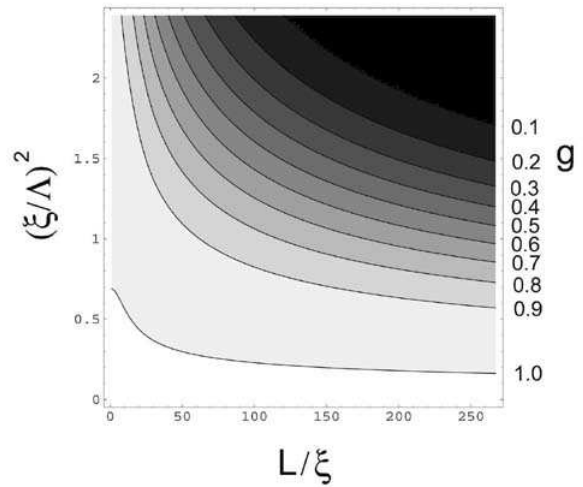}
\end{center}
\caption[a]{\narrowtext The contour plot corresponding to
Fig.~\ref{fig3}. The value of $g$ along each contour line is
indicated on the right vertical axis.} \label{fig4}
\end{figure}
This displays clearly the crossover from a condensate near the
origin and near the axes, where $g \simeq 1$, to a quasicondensate
in the regions away from the origin and the axes. Fig.~\ref{fig4}
is a two-dimensional contour plot that is equivalent to
Fig.~\ref{fig3}. The contour lines in this figure show the
relation between $L/\xi$ and $(\xi/\Lambda)^2$ for a fixed value
of $g$. Note that when quantum fluctuations can be neglected, we
expect that $-\ln [g(0.9L)] \propto (L/N_0\xi)(\xi/\Lambda)^2
(0.9L/\xi)$, which shows that for a fixed value of $g$ we have
that $(\xi/\Lambda)^2 \propto (L/\xi)^{-2}$. This relation is
indeed borne out by a detailed inspection of Fig.~\ref{fig4}. The
theory discussed in this paper is only valid in the
weakly-interacting limit $4\pi n\xi\gg1$ and therefore does not
include the Tonks regime \cite{{tonks1},{tonks2}}. Nonetheless, it
is instructive to indicate at what region of the phase diagram
such a regime may exist. The criteria for the Tonks gas limit is
$N<4\pi\kappa L$ \cite{2d,maxim}. Upon using the above-mentioned
scaling, this limit reads $L/\xi>\sqrt{2}N$. Since Fig.~\ref{fig3}
was generated with $N=2000$, it is clear that in order to be in
the Tonks gas limit, the condition $L/\xi>2800$ must be satisfied.

\subsection{Trapped Bose gas}

Next, we construct the phase diagram of a one-dimensional Bose gas
confined by a harmonic potential, in order to investigate the
effect of the induced density inhomogeneity. Our treatment is
based on the local-density approximation for the calculation of
the density profile, and is performed within the many-body
T-matrix approximation for the interatomic interactions (as
opposed to the two-body limit used in the preceeding homogeneous
calculations). The calculation of the correlation function is
again performed for a fixed number of atoms, but using now
Eq.~(\ref{chifinal}). In the trapped case, a measure of the size
of the system can be determined by the
 Thomas-Fermi radius $R_{\rm TF}$,
which is, in general, a function of both temperature and trapping
frequency. To plot a universal phase diagram, however, the axis
corresponding to the system size should be independent of the
other axis representing the temperature. This argument is
strengthened by the fact that the temperature-dependent
Thomas-Fermi radius $R_{\rm TF}(T)$ does not increase
monotonically with increasing confinement at fixed temperature and
total number of atoms. Such an effect is a direct consequence of
two competing mechanisms which have a different dependence on the
confinement: As the trap frequency increases, one would in first
instance expect the size of the (quasi)condensate to decrease
monotonically. However, due to the constraint on the total number
of atoms, the number of atoms in the (quasi)condensate tends to
increase simultaneously. This can lead to an increase in $R_{\rm
TF}(T)$, in particular at higher temperatures.
\begin{figure}[htb]
\begin{center}
\includegraphics[width=8cm]{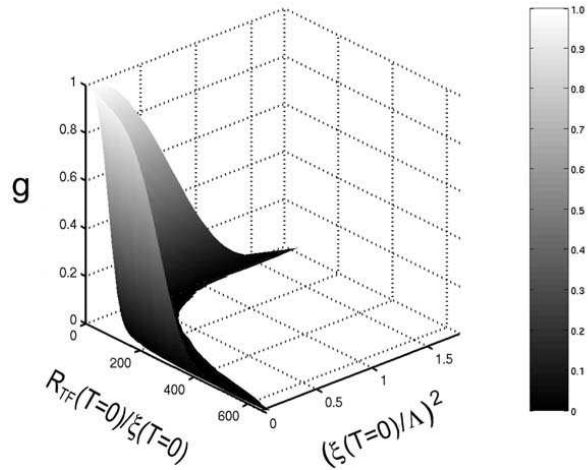}
\end{center}
\caption[a]{
Universal phase diagram of a pure one-dimensional trapped atomic
Bose gas with constant number of atoms.
 Greyscale indicates the value of $g(0,z=0.9 R_{TF}(T))$
with lighter regions corresponding to maximum coherence $g \simeq
1$. This figure probes the entire region from the
normal-to-(quasi)condensate crossover (where the (quasi)condensate
depletion approaches 100\%, curved region away from origin), to
the presence of true Bose-Einstein condensation, for which the
quasi-condensate depletion becomes negligible near the origin. The
entire phase diagram plotted here remains well within the validity
of our Thomas-Fermi approximation, with $\gamma = N_{0} \kappa
l_{z} \gg 1$ and $N_{0}$ the number of (quasi)condensate atoms.
For a $^{23}Na$ condensate with $N=2000$ atoms under
 very tight transverse confinement of
$\omega_{\perp}/2 \pi = 5$ KHz, the above phase diagram would
correspond to temperatures in the range $0.2 nK - 12 \mu K$ and
longitudinal frequencies $\omega_{z} /2 \pi$ from 0.05 to 45000
Hz. } \label{fig5}
\end{figure}

To obtain a universal phase diagram in the case of a fixed total
number of atoms, we hence use the zero-temperature Thomas-Fermi
radius  $R_{\rm TF}(T=0)$ as a measure of the size of the system.
This is again scaled to the zero-temperature correlation length
$\xi(T=0)$. The latter is evaluated in terms of the central
density $n(0)$ via $\xi(T=0)=1/\sqrt{8 \pi \kappa n(0)} = \hbar
/\sqrt{2 \pi \mu m}$ where the one-dimensional chemical potential
at zero temperature in the Thomas-Fermi approximation is given by
$\mu = (3 \pi /\sqrt{2})^{2/3} (N \kappa l_{z})^{2/3} \hbar
\omega_{z}$ \cite{jhun}. Note that the actual size of the
(quasi)condensate is given by the temperature-dependent
$R_{TF}(T)$, so a good measure of the macroscopic coherence of the
system is obtained by measuring the correlation function at $0.9
R_{TF}(T)$ \cite{jhun}.

\begin{figure}[htb]
\begin{center}
\begin{math}
\begin{array}{cc}
\includegraphics[width=7.cm]{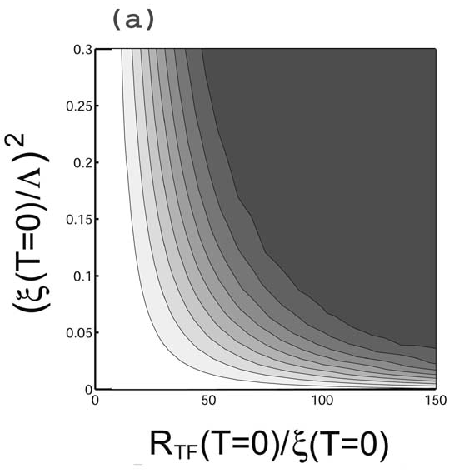}
&\\
\includegraphics[width=7.cm]{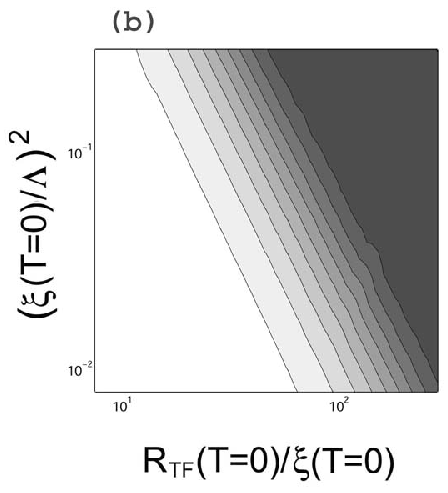}
&
\end{array}
\end{math}
\end{center}
\caption[a]{
Contour plot corresponding to Fig.~\ref{fig5}, with both axes
plotted on (a) linear and (b) logarithmic scale. As in
Fig.~\ref{fig4}, each contour corresponds to a value of $g$
ranging from $0.1$ (leftmost contour) to $0.9$ (rightmost
contour). White regions correspond to true Bose-Einstein
condensation, and darkest regions to the absence of phase
coherence over the system size. Note that greyscale in this figure
is not the same as in Fig.~\ref{fig5}. Note also that we have not
explicitly computed for $R_{\rm TF}(T=0) / \xi(T=0) < 7$, where
there is essentially true Bose-Einstein condensation, as this
requires extremely high temperatures and large longitudinal
trapping frequencies.
 The condition of validity of the Thomas-Fermi
approximation will also break down in this limit. } \label{fig6}
\end{figure}

Fig.~\ref{fig5} shows the value of the normalized correlation
function of Eq.~(\ref{d11}) evaluated at $z = 0.9 R_{TF}(T)$ as a
function of $R_{TF}(T=0)/\xi(T=0)$ and $(\xi(T=0)/\Lambda)^{2}$
for fixed coupling constant $\kappa$ and total number of atoms $N$
in the system. Light color indicates the presence of a pure
condensate and black color indicates the absence of universal
phase coherence across the system. As expected, the coherence is
found to increase when decreasing the size of the system, i.e.,
increasing the trap frequency $\omega_{z}$ at fixed temperature,
or decreasing the temperature in a fixed trap. This is similar to
the homogeneous case, with $g$ becoming close to one as
$R_{TF}(T=0)/\xi(T=0)$ or $(\xi(T=0)/\Lambda)^{2}$ become small.
Note that the quasicondensate depletion due to interactions and
nonzero temperatures can be very significant, and this figure
covers the entire range of (quasi)condensate fractions, ranging
from one at the origin, where we essentially deal with a pure
condensate, to approximately zero near the transition to the
normal state sufficiently far from the origin. In particular, the
characteristic curved shape away from the origin marks a sharp
transition from a quasi-condensate to the normal phase. Note that
we find a sharp transition due to the use of the local-density
approximation, whereas in reality this would be a smooth
crossover. To study now the crossover between regions of different
degree of coherence within the (quasi)condensate regime,
Fig.~\ref{fig6}(a) shows the corresponding contour plot, focusing
on a small region near the origin, where the crossover takes
place. In this figure, white regions correspond to true
Bose-Einstein condensation, whereas the darkest region on the top
right corner of the figure corresponds to complete absence of
phase coherence across the size of the system. The corresponding
contour lines with both axes plotted on logarithmic scale is shown
in Fig.~\ref{fig6}(b).

%
The above phase diagram corresponds to the limit of a pure 1D Bose
gas. In trying to probe this experimentally by means of the
recently produced quasi-1D Bose gases
\cite{lowdketterle,exp100,exp200,exp300,exp500}, there are certain
constraints which must be taken into consideration. First, the
temperature must be low enough that transverse thermal excitations
are suppressed, i.e., $ k T < \hbar \omega_{\perp}$.
 Second, the chemical potential must not exceed
the transverse energy, i.e., $\mu < \hbar \omega_{\perp}$ thus
restricting the maximum longitudinal frequencies for which the
system remains kinematically one dimensional, to some fraction of
the transverse confinement. The exact fraction depends on the
atomic species, the scattering length, the number of atoms and the
temperature. Finally, the longitudinal confinement cannot be made
arbitrarily weak, as it will then be essentially impossible to
trap the atoms. All above constraints require a very large
transverse confinement, since the experimentally-accessible region
of such a phase diagram increases with increasing transverse
confinement. The experimental constraint of minimum longitudinal
frequency does not pose much of a limitation since the low
frequency regions correspond to the outmost points of
Fig.~\ref{fig5}, for which the macroscopic coherence has already
been completely lost. Of the other two constraints, which is more
restrictive depends on the details of the particular experiment,
since they exclude different regions of the phase diagram.

\begin{figure}[htb]
\begin{center}
\begin{math}
\begin{array}{cc}
\includegraphics[width=7.cm]{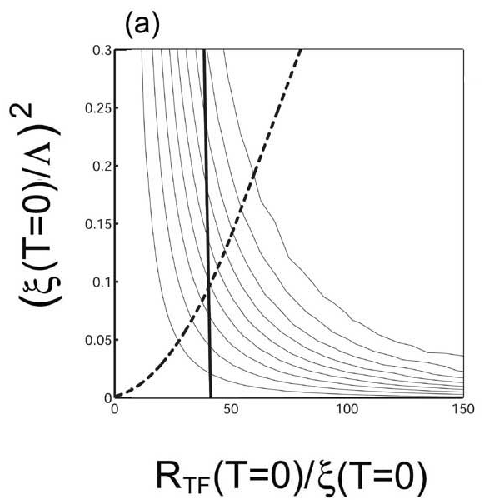}
&\\
\includegraphics[width=7.cm]{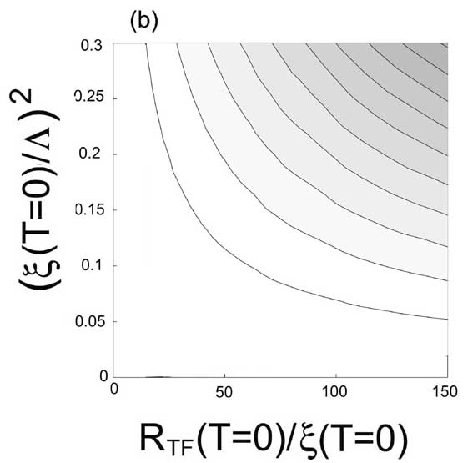}
&
\end{array}
\end{math}
\end{center}
\caption[a]{
(a) Same contour plot as in Fig.~\ref{fig6}(a) indicating clearly
the regions experimentally accessible for $N=2,000$ $^{23}Na$
atoms with $\omega_{\perp}/2 \pi = 5$ KHz. The solid line
corresponds to the condition $\mu = \hbar \omega_{\perp}$, whereas
the dashed line indicates the regime where $ k T = \hbar
\omega_{\perp}$.
 Experimentally accessible regions due to
actual quasi-1D constraints are located to the right of each of
these lines. (b) Corresponding contour plot of quasi-condensate
fraction. Each contour corresponds to a change in $N_0/N = 0.05$,
with the leftmost contour indicating the region where
$N_0/N=0.95$, and the rightmost contour in the plotted region
corresponding to $N_0/N=0.5$. It is thus evident that the
quasi-condensate depletion becomes significant very rapidly, as
one moves away from the origin of the figure. } \label{fig7}
\end{figure}
To illustrate this, Fig.~\ref{fig7}(a) indicates the
experimentally accessible regions for the case of $N=2000$
$^{23}Na$ condensate atoms under a transverse confinement of
$\omega_{\perp}/2 \pi = 5$ KHz \cite{Hannover}.
 The
chemical potential constraint generally excludes a region to the
left of Fig.~\ref{fig6}(a), and so large coherences are best
probed by restricting the condensate to a very small number of
atoms. For the case studied here, this translates to the
approximate condition $R_{\rm TF}(T=0) / \xi(T=0) > 40$. The
temperature constraint on the other hand prohibits a large part of
the top left of Fig.~\ref{fig6}(a), which requires the experiments
to be performed at sufficiently low temperatures (here $T << 240
nK$). We see that for low enough temperatures and small enough
atom numbers, it is possible to probe experimentally the entire
phase coherence regime (from $g=0$ to roughly $g=0.99$). Although
the coherence decreases with increasing scattering length, for
example by tuning around a Feshbach resonance \cite{ket55,car55},
or central density, this can further restrict the region of
accessibility due to the chemical potential constraint.

Fig.7(b) gives the corresponding contour plot of the
(quasi)condensate fraction. It is evident that the
(quasi)condensate depletion becomes important even within the
range of this figure, with the furthermost points here
corresponding to more than $50\%$ depletion. In our analysis, we
have chosen to monitor the loss of phase coherence by looking at
the amount of decrease of the normalized two-point
 correlation function near the edge of the system (at $z=0.9R_{TF}$). However, it
may be experimentally easier to look at the decay of this
correlation function at a point somewhat closer to the centre.
This would then lead to a slower decay of phase coherence across
the measured distance, and a sufficiently small amount of phase
coherence across such a region would then occur for even larger
(quasi)condensate depletions. Therefore, we believe that a
quantitative study of the phase-coherence crossover problem
requires a theory which takes into account this depletion.

\section{Conclusions}
\label{conc}

We have obtained a detailed universal picture of the coherence
properties of a Bose gas as a function of the dimensionality, the
temperature and the interaction strength, paying particular
attention to the crossover from quasicondensation to true
Bose-Einstein condensation. In a finite homogeneous system at zero
temperature, the long-range coherence is lost when the dimension
of the system is reduced to one dimension. This dimensional
reduction leads to the appearance of a quasicondensate, whose
coherence
 extends over a range that is much larger than the coherence
length but much smaller than the size of the system. Similar
results were obtained when the effects of temperature and the
strength of the mean-field interaction were investigated. In a
homogeneous one-dimensional system, a true condensate is located
in a region in the phase diagram where both the length of the
system and its temperature are small. Away from this region, a
quasicondensate is present covering a wider region of the phase
diagram.

A similar phase diagram was obtained for a harmonically confined
system in one dimension, where a change in the confining potential
at fixed temperature leads to a change in the size of the
(quasi)condensate. It is important to note that our theory covers
the entire temperature range, from zero temperature up to the
crossover to the normal phase, where the (quasi)condensate becomes
fully depleted. Since our work takes the (quasi)condensate
depletion into account, the presented results extend beyond
existing treatments which have been limited to the regime of
negligible depletion \cite{1d,2d,petrov22}. Note also that the
phase diagram presented in Sec. IV corresponds to the pure
one-dimensional limit. Nonetheless, a large part of this, ranging
from, complete absence of, to complete coherence over the system,
can be probed experimentally with existing technology, as
discussed by a particular example based on typical parameters.

\section*{Acknowledgments}
This work was supported by the Stichting voor
Fundamenteel Onderzoek der Materie
(FOM), which is supported by the Nederlandse Organisatie voor Wetenschappelijk
Onderzoek (NWO).
N.P.P. acknowledges funding from the U.K. EPSRC and the hospitality of the ITP.


\end{document}